\documentclass[a4paper,11pt]{article}
\usepackage{jcappub} % for details on the use of the package, please
\usepackage[T1]{fontenc} % if needed
\DeclareUnicodeCharacter{2212}{\ensuremath{-}}
\DeclareUnicodeCharacter{2260}{\ensuremath{-}}

\title{\boldmath A study of the pulsar EXO 1745-248 in $f(Q)$ gravity with pseudo-spheroidal geometry  }

\author[a]{Bibhash Das\footnote{ORCID : 0000-0001-7283-6745}}
\author[a]{Bikash Chandra Paul\footnote{ORCID : 0000-0001-5675-5857}}

\affiliation[a]{Department of Physics, University of North Bengal, Darjeeling, West Bengal, India -734 013}
\emailAdd{rs\_bibhash@nbu.ac.in}
\emailAdd{bcpaul@nbu.ac.in (corresponding author)}
\abstract{ We present a singularity-free relativistic interior solution for constructing stable quark stellar models in the framework of a linear $f(Q)$ gravity ($f(Q) = \alpha Q + \phi$) satisfying the pseudo-spheroidal geometry. The physical features and the stability of the stellar model is explored with strange star (SS) candidate EXO 1745-248 ($M = 1.7\, M_{\odot}$ and $R = 9\, km$). The Durgapal-Banerjee transformation is employed to obtain the relativistic interior solution using the MIT Bag model equation of state (EoS): $P = \frac{1}{3}(\rho - 4 B_{g})$. For a linear form of $f(Q)$ gravity, we obtain the exterior vacuum solution, which reduces to the Schwarzschild-de Sitter (SdS) solution with the cosmological constant term, $\Lambda = \frac{\phi}{2\alpha}$. The stellar model is analyzed for the different values of the spheroidicity parameter ($\mu$). The value of $\alpha$ is constrained using a viable physical limit on the Bag parameter ($B_{g} \in [57.55,95.11]\,MeV\,fm^{-3}$).  The constraints on Mass-Radius relation indicates that physically acceptable SS models are permitted for $\mu \geq 7$. The contribution of $\mu$ to the energy density, pressure profiles, and other physical features is studied for the SS candidate EXO 1745-248. The stability of the stellar model obtained here is also analyzed through causality condition, adiabatic index and other stability criteria. We also investigate the stellar model for other SS candidates to test its viability. The relativistic interior solution obtained here can be used to construct viable and physically acceptable strange star models with very high compactness ratio in the framework of linear $f(Q)$ gravity. 
}

\begin{document}
\maketitle
\flushbottom

\section{Introduction}
\label{intro}
Einstein’s theory of General Relativity (GR) \cite{Einstein1915OnRelativity} has stood for over a century as the foundation of modern gravitational physics, successfully explaining phenomena from planetary motion to the bending of light around massive bodies. Numerous experimental and observational tests have been conducted in both astrophysical and cosmological contexts to evaluate the validity of General Relativity (GR) \cite{Will2014TheExperiment,Ishak2019TestingCosmology}. Despite its success, GR encounters significant shortcomings when it comes to topics like the late-time cosmic acceleration of the universe, the dark matter and dark energy problems, and the existence of spacetime singularities in the Big Bang and inside Black Holes (BHs). Now, there are two methodologies to address these issues: the first involves altering the matter sector by incorporating additional `dark' components into the universe's energy budget, while the second entails modifying the geometrical part of GR. In the second approach, the modifications are made by extending the Einstein–Hilbert action in an attempt to capture new geometric or dynamical aspects of gravity beyond GR. 

The standard description of GR makes use of the Riemannian geometry by specifying a metric-compatible affine connection on the spacetime manifold, such as the Levi-Civita connection. Within the context of Riemannian geometry, GR can be extended by substituting the Ricci scalar ($R$) on the Einstein-Hilbert action by a function of $R$, or a function with a combination of $R$ and $\mathcal{T}$, where $\mathcal{T}$ is the trace of the energy-momentum tensor. These extensions of GR based on the curvature are termed as $f(R)$ and $f(R,\mathcal{T})$ theories of gravity. However, there can be a variety of choices of the affine connection on any manifold, and these connections may represent different but equivalent descriptions of gravity with more general geometric structures. In his attempt to unify electromagnetism and gravitation, Weyl \cite{Weyl1918GravitationElectricity} replaced the metric field by a class of all conformally equivalent metrics and introduced an extra connection that does not contain any information about the length of a vector in parallel transport. In this theory, the covariant divergence of the metric tensor is non-zero, which can be expressed mathematically through a new geometric quantity known as non-metricity ($Q$). Similarly, Cartan's work \cite{Cartan1922OnSpaces} was another significant advancement in geometry that gave rise to a new class of generalized geometric theories of gravity by introducing a new geometric element known as the torsion ($T$). This new generalized geometric gravity theory was proposed as an extension of GR \cite{Cartan1923OnPart,Cartan1924OnOnecontinued,Cartan1925OnTwo}, today known as the Einstein-Cartan theory \cite{Hehl1976GeneralProspects}. The torsion can be naturally included in the extended Weyl geometry, which results in the Weyl-Cartan geometry. Based on these new geometric quantities, GR can be modified with non-Riemannian geometry by two equivalent geometric extensions : (i) extensions based on torsion ($T$), where curvature and non-metricity ($Q$) vanish (\textit{viz.} $f(T)$ gravity), and (ii) extensions based on non-metricity ($Q$), where the curvature and torsion ($T$) vanish (\textit{viz.} $f(Q)$ gravity). These theories of gravity that are based on the non-Riemannian geometry are known as the Teleparallel Equivalent of General Relativity (TEGR) \cite{Hayashi1979NewRelativity,Sauer2006FieldTheory} and the Symmetric Teleparallel Equivalent of General Relativity (STEGR) \cite{Nester1999SymmetricRelativity,Jimenez2018CoincidentRelativity,Jimenez2018TeleparallelTheories}, respectively. Under the teleparallelism and torsion-free constraints, a specific gauge, known as the `coincident gauge' can be used to formulate this theory, in which the covariant derivatives are reduced to partial derivatives. However, using this gauge can cause the metric to evolve differently in different coordinate systems. In the present work, we have considered a non-coincident gauge following the work of Alwan et.al. \cite{Alwan2024}.
%TEGR can be formulated by choosing a connection where both the curvature ($R$) and non-metricity ($Q$) vanish while the torsion ($T$) is non-zero. On the other hand, a flat spacetime manifold with non-vanishing nonmetricity ($Q$) but no curvature ($R$) and torsion ($T$) gives rise to the STEGR, where $Q$ facilitates gravitational interaction.

The Symmetric Teleparallel Equivalent of General Relativity (STEGR) has developed into the coincident gravity theory, commonly referred to as the $f(Q)$ gravity theory, which bears a notable similarity to the $f(R)$ theory \cite{Heisenberg2019AImplications}. Recent years have seen numerous studies on $f(Q)$ gravity theory and its astrophysical and cosmological implications, including late-time acceleration \cite{Lazkoz2019ObservationalGravity},  bouncing cosmology \cite{Bajardi2020BouncingGravity}, and several other aspects of $f(Q)$ gravity in cosmology \cite{Harko2018CouplingGravity,Frusciante2021SignaturesCosmology,Narawade2022DynamicalGravity}. A thorough analysis of $f(Q)$ gravity is provided in Ref. \cite{Heisenberg2024ReviewGravity} and the references within. Aside from its application in cosmology, $f(Q)$ gravity has been extensively applied in the study of compact astrophysical entities, including black holes (BHs). \cite{DAmbrosio2022BlackGravity,Calza2023AFQ-gravity}. Apart from BHs, neutron stars also serve as natural laboratories for studying the modified theories of gravity due to their extreme densities and strong gravitational fields. The additional degrees of freedom, scalar fields, or the geometric corrections introduced in these modified theories of gravity can influence the stability, hydrostatic equilibrium, and the mass-radius relation of a neutron star. Lin et.al. \cite{Lin2021SphericallyGravity} explored the effects of $f(Q)$ by considering internal and external solutions in a spherically symmetric configuration. Using gravitational decoupling, Maurya et.al. \cite{Maurya2023TheTheory} constructed stable strange stars in the framework of $f(Q)$ gravity with a quadratic equation of state. Bhattacharjee et al. \cite{Bhattacharjee2025Charged-action} constructed a charged analogue of an anisotropic star under linear $f(Q)$ gravity employing the Krori-Barua metric ansatz. A number of recent studies on $f(Q)$ gravity are available in the literature \cite{Mandal2022AField,Sokoliuk2022BuchdahlTheory,Maurya2022ExploringGravity,Bhar2023PhysicalGravity,Ditta2023AnisotropicGravity,Dimakis2024StaticTheory,Sharif2024ImpactGravity,Bhattacharjee2025InteractingStars}. 

A compact star is assumed to be a perfectly uniform, or isotropic sphere, which is often an oversimplification. In the extreme conditions found in these objects, a phenomenon known as the pressure anisotropy can occur, which was first pointed out by Canuto \cite{Canuto1974EquationDensities}. This means the pressure pushing outward from the star's center, known as the radial pressure ($P_r$), is not the same as the pressure exerted parallel to the surface, called the tangential pressure ($P_{\perp}$). This directional difference in pressure becomes almost inevitable when matter densities reach immense levels of about $\sim 10^{15} \,gm\,cm^{-3}$ \cite{Ruderman1972Pulsars:Dynamics}. Several physical processes are believed to cause this state \cite{Herrera1997LocalSystems}. The presence of a star's incredibly strong magnetic field \cite{Weber2017PulsarsPhysics} is a primary contributor , but other factors like phase transitions \cite{Sokolov1980PhaseLiquid}, pion condensation \cite{Sawyer1972CondensedMatter}, the shear of the stellar fluid \cite{DiPrisco2007NonadiabaticCollapse}, and even the formation of a solid core \cite{Kippenhahn1990StellarEvolution} can also lead to anisotropy. This difference in pressure creates an extra force inside the star, which has a significant effect on its hydrostatic equilibrium and overall stability, which makes studying it necessary for making accurate models of stars. Recently, numerous studies have emerged that investigate the impact of pressure anisotropy within the context of GR and modified gravitational theories. Maurya et.al. \cite{Maurya2017RelativisticF} obtained an exact solution with anisotropic matter distribution in Buchdhal-type relativistic stars. Ratanpal et al. \cite{Ratanpal2023AnisotropicModel} have taken an anisotropic approach to construct stable stellar models of compact stars in GR. A new class of relativistic anisotropic stellar model in higher-dimensional Einstein-Gauss-Bonnet gravity is obtained by Das et.al. \cite{Das2022}. Further, spherically symmetric charged anisotropic stars have been studied in GR and in modified Rastall gravity by Das et.al. \cite{Das2023ModelsMetric} and Bhattacharjee et.al. \cite{Bhattacharjee2025Singularity-freeGravity}, respectively. Similar studies on anisotropy can be found in the Refs. \cite{Maurya2018RoleStars,Chanda2019AnisotropicGeometry,Dey2020HigherGeometry,Dey2021CompactGeometry,Das2023Anisotropic0,Das2024Finch-SkeaAnisotropy,Kaur2024ChargedMetric}. 

In order to construct a stable stellar configuration of a compact star, the spacetime geometry must be provided to solve the field equations in GR or in modified gravity. Several geometries have been suggested in this direction, namely, Generalized Tolman-Kuchowicz (GTK) metric \cite{Das2023ModelsMetric}, Krori-Barua (KB) metric \cite{Krori1975ARelativity}, Finch-Skea (FS) metric \cite{Finch1989ARay}, Pseudo-spheroidal metric \cite{Tikekar1998RelativisticSpace-time}, etc. In this text, the radial component ($e^{\lambda(r)}$) is provided with the the pseudo-spheroidal metric ansatz to describe a part of the interior spacetime of the star. Several works have been done to study compact stars satisfying the pseudo-spheroidal metric \cite{TIKEKAR2005RELATIVISTICSPACETIME,Chattopadhyay2010RelativisticSpace-time,CHATTOPADHYAY2012RELATIVISTICSPACETIME,Ratanpal2016AnisotropicSpacetime,Shee2017CompactSpacetime}. Now, to derive the temporal component $e^{\nu(r)}$ of the interior solution using the field equations, it is necessary to consider an Equation of State (EoS) that describes the internal composition of compact stars and gives a relationship between density and pressure (considering the interior of the compact star is isothermal). Researchers have shown significant interest in examining the internal composition of compact stars from various perspectives over the years. One such promising theory is the Quark star hypothesis. In this direction, the MIT Bag model EoS \cite{Chodos1974NewHadrons} has gained a bit of popularity, describing how bulk quark matter behaves and its deconfined state within a spatial region referred to as a `Bag.' The EoS is given by, $P = \frac{1}{3}(\rho - 4 B_{g})$, where $P$ is the pressure, $\rho$ is the energy density, and $B_{g}$ is the Bag constant. The MIT Bag model treats quarks as a degenerate Fermi gas composed of $u$, $d$, massive $s$ quarks and electrons. Madsen \cite{Madsen1998PhysicsMatter} obtained a limit on the Bag constant, $ B_{g}\in [57.55,95.11]\,MeV\,fm^{-3}$ for stable SQM. This triggered an investigation on an entirely new class of stars that are composed of strange quark matter (SQM) and the said stars are called Strange Quark Stars (SQS) or Strange Stars (SS). EXO 1745-248, Her X-1, PSR J1903+0327, PSR J1614-2230, 4U 1820-30, etc., are some of the candidates for strange stars. 

The motivation of the paper is to obtain a singularity-free relativistic interior solution for obtaining a stable strange star model in the framework of a linear $f(Q)$ gravity satisfying pseudo-spheroidal geometry and study the physical features and the stability of the SS candidate EXO 1745-248 \cite{Ozel2009THE1745248}. The paper is presented as follows: Section \ref{sec:2} presents the derivation of the basic mathematical formulation in $f(Q)$ gravity. The field equations in the linear $f(Q)$ gravity are derived in Section \ref{sec:3}. In Section \ref{sec:4}, we obtained the exact solutions of the field equations in pseudo-spheroidal geometry employing the Durgapal-Banerjee transformation. The exterior is obtained as the boundary conditions are discussed in Section \ref{sec:5}. Section \ref{sec:6} discussed the minimum criteria for obtaining a realistic, stable stellar model. We physically analyze different physical properties, i.e., energy density profile, pressure profile, anisotropy, energy conditions, etc., in Section \ref{sec:7}. The stability of the stellar model is explored in Section \ref{sec:8}. In Section \ref{sec:9}, we physically analyze a few other SS candidates. Section \ref{sec:10} finally presents a brief discussion.

%In the extensions based on curvature, the Levi-Civita connection requires that $R$ remain non-zero, while $T$ and $Q$, should both vanish. By relaxing these constraints, one can formulate theories of gravity  based on the non-Riemannian. In the later case, the extensions based on torsion ($T$) and non-metricity ($Q$)

%%%%%%%%%%%%%%%%%%%%%%%%%%%%%%%%%%%%%%%%%%%%%%%%%%%%%%%%%%%%%%%%%%%%%%%%%

\section{Basic mathematical formalism in $f(Q)$ gravity}
\label{sec:2}
The affine connection in the case of Weyl-Cartan geometry can be decomposed into three irreducible parts, i.e., the Levi-Civita connection ($\{ ^{\, \chi}_{\eta \kappa} \}$), the contortion tensor ($K^{\chi}_{\eta \kappa}$), and the disformation tensor ($L^{\chi}_{\eta \kappa}$) as follows
    \begin{equation}
        \label{ac}
        \Gamma^{\chi}_{\eta \kappa} = \{ ^{\,\chi}_{\eta \kappa} \} + K^{\chi}_{\eta \kappa} + L^{\chi}_{\eta \kappa},
    \end{equation}
where, $\{ ^{\chi}_{\eta \kappa} \} = \frac{1}{2}g^{\chi \beta}\,(\partial_{\eta}g_{\beta \kappa} + \partial_{\kappa}g_{\beta \eta} - \partial_{\beta}g_{\eta \kappa})$, $K^{\chi}_{\eta \kappa} = \frac{1}{2}\left( T^{\chi}_{\eta \kappa} + T_{\eta}\, ^{\chi}_{\kappa} + T_{\kappa}\, ^{\chi}_{\eta} \right)$, and $L^{\chi}_{\eta \kappa} = \frac{1}{2}\left( Q^{\chi}_{\eta \kappa} - Q_{\eta}\, ^{\chi}_{\kappa} - Q_{\kappa}\, ^{\chi}_{\eta} \right)$.\\

The expression for the non-metricity tensor ($Q_{\chi \eta \kappa}$) in terms of the metric tensor $g_{\eta \kappa}$ and the affine connection $\Gamma^{\chi}_{\eta \kappa}$ is given by
    \begin{equation}
        Q_{\chi \eta \kappa} := \nabla_{\chi} g_{\eta \kappa} = \partial_{\chi} g_{\eta \kappa} - \Gamma ^{\beta}_{\chi \eta} g_{\beta \kappa} - \Gamma ^{\beta}_{\chi \kappa} g_{\beta \eta}
    \end{equation}
and the torsion tensor ($T^{\chi}_{\eta \kappa}$) is given by
    \begin{equation}
        T^{\chi}_{\eta \kappa} := \Gamma^{\chi}_{\eta \kappa} - \Gamma^{\chi}_{\kappa \eta}.
    \end{equation}
The non-metricity scalar is given by
    \begin{equation}
        \label{nonmetricity}
        Q = Q_{\chi \eta \kappa}\, P^{\chi \eta \kappa}. 
    \end{equation}
Here $P^{\alpha \beta \gamma}$ is the non-metricity conjugate, which has the following form:
    \begin{equation}
        P^{\chi}_{\eta \kappa} = -\frac{1}{4} Q^{\chi}_{\eta \kappa} + \frac{1}{4} \left( Q_{\eta}\, ^{\chi}_{\kappa} + Q_{\kappa}\, ^{\chi}_{\eta} \right) + \frac{1}{4}Q^{\chi}g_{\eta \kappa} - \frac{1}{8}\left( 2 \Tilde{Q}^{\chi}g_{\eta \kappa} + \delta^{\chi}_{\eta} Q_{\kappa} + \delta^{\chi}_{\kappa} Q_{\eta} \right).
    \end{equation}
The components in the affine connection in Eq. (\ref{ac}) can be rewritten as,
    \begin{equation}
        \Gamma^{\chi}_{\eta \beta} = \frac{\partial \varphi^{\chi}}{\partial \xi^{\sigma}}\partial_{\eta} \partial_{\beta} \xi^{\sigma}.
    \end{equation}
In the above equation, $\xi^{\chi} = \xi^{\chi}(\varphi^{\eta})$ is an invertible relation and $\frac{\partial \varphi^{\chi}}{\partial \xi^{\sigma}}$ is the inverse of the corresponding Jacobian \cite{Jimnez2021}. This is the case known as a coincident gauge, where there is a chance of getting a coordinate system for which, $\Gamma^{\chi}_{\eta \kappa} = 0$. Consequently, the covariant derivative ($\nabla_{\chi}$) reduces to the partial derivative ($\partial_{\chi}$). Thus the Levi-Citiva connection can be expressed in terms of the disformation tensor as, $\{ ^{\,\chi}_{\eta \kappa} \} = - L^{\chi}_{\eta \kappa}$.  

The action in $f(Q)$ gravity is  given by,
    \begin{equation}
        \label{actionterm}
        S = \int \Big[ f(Q) + \mathcal{L}_M \Big]\sqrt{-g}\, d^4 x
    \end{equation}
here, $f(Q)$ representing a general function of the non-metricity $Q$ and $\mathcal{L}_M$ denoting the Lagrangian density of the matter fields. Now by varying the action term in Eq. (\ref{actionterm}) with respect to the metric tensor $g_{\eta \kappa}$, the field equations for the $f(Q)$ gravity can be obtained as
    \begin{equation}
        \label{varyaction1}
        \frac{2}{\sqrt{-g}} \nabla_{\chi} \left( \sqrt{-g}\,f_{Q}\,P^{\chi}_{\eta \kappa} \right) + \frac{1}{2} g_{\eta \kappa}\,f + f_{Q} \left( P_{\eta \chi i}\,Q_{\kappa}^{\chi \sigma} - 2 Q_{\chi \sigma \eta} P^{\chi \sigma}_{\kappa} \right) = - 8\pi \mathcal{T}_{\eta \kappa}
    \end{equation}
where $f_Q \equiv \frac{df}{dQ}$ and the energy-momentum tensor ($\mathcal{T}_{\eta \kappa}$) takes the form
    \begin{equation}
        \label{energymomentum}
        \mathcal{T}_{\eta \kappa} = -\frac{2}{\sqrt{-g}} \frac{\delta \left(\sqrt{-g}\, \mathcal{L}_{M} \right)}{\delta g^{\eta \kappa}}.
    \end{equation}
We also obtain the following equation by varying the action term in Eq. (\ref{actionterm}) with respect to the affine connection:
    \begin{equation}
        \label{varyaction2}
        \nabla_{\eta} \nabla_{\kappa} \left( \sqrt{-g}\, f_{Q}\, P^{\eta \kappa}_{\chi} \right) = 0
    \end{equation}
    It is also noted that the field equation guarantees the conservation of the energy-momentum tensor and the Einstein's equation can be recovered by choosing $f(Q)= Q$.
%Now with the help of the last three Eqs (\ref{varyaction1}) - (\ref{varyaction2}), we will derive the exact form of the field Eqs in the following section.

%%%%%%%%%%%%%%%%%%%%%%%%%%%%%%%%%%%%%%%%%%%%%%%%%%%%%%%%%%%%%%%%%%%%%%%%%

\section{Field equations in $f(Q)$ gravity}
\label{sec:3}
For a spherically symmetric system, we consider the following line element in a curvature coordinate system,
    \begin{equation}
        \label{lineelement}
        ds^2 = -e^{\nu(r)}\, dt^2 + e^{\lambda(r)}\,dr^2 + r^2 \left( d\theta^2 + \sin^2\theta d\phi^2 \right)
    \end{equation}
where $\nu(r)$ and $\lambda(r)$ are the unknown metric potentials and are functions of the radial coordinate $r$. The metric potentials $\nu(r)$ and $\lambda(r)$ tend to $0$ as $r$ $\to$ $\infty$, the spacetime will be asymptotically flat. For perfect fluid matter distribution, the energy-momentum tensor is given by,
    \begin{equation}
        \label{perflu}
         \mathcal{T}_{\eta \kappa} = diag (- \rho, P_{r}, P_{\perp}, P_{\perp})
    \end{equation}
where, $\rho$ is the energy-density, $P_r$ is the radial pressure, and $P_{\perp}$ is the tangential pressure. Considering the non-coincident gauge case, i.e., $\Gamma^{\chi}_{\eta \kappa} \neq 0$ and using Eqs. (\ref{ac}), (\ref{lineelement}), and (\ref{perflu}), the non-vanishing components of the connections can be obtained as
    \begin{equation}
        \Gamma^{\theta}_{r\theta} = \Gamma^{\theta}_{\theta r} = \Gamma^{\phi}_{r \phi} = \Gamma^{\phi}_{\phi r} = \frac{1}{r}; \,\,\,\,\,\,\,\,\,\,\,\, \Gamma^{r}_{\theta \theta} = -r, \nonumber
    \end{equation}
    \begin{equation}
        \label{nvac}
        \Gamma^{\phi}_{\theta \phi} = \Gamma^{\phi}_{\phi \theta} = \cot \theta;\,\,\,\,\,\,\,\, \Gamma^{r}_{\phi \phi} = -r \sin^2 \theta, \,\,\,\,\,\,\,\, \Gamma^{\theta}_{\phi \phi} = -\cos\theta \sin\theta.
    \end{equation}
For static spherically symmetric spacetime in $f(Q)$ theory, we use the affine connection given in Eq. (\ref{nvac}) and the field equations (FE) from Eq. (\ref{varyaction1}) in the framework of $f(Q)$ gravity yield \cite{Alwan2024},
    \begin{multline}
       \label{rho1}
        8\pi \rho(r) = \frac{1}{2r^2\ e^{\lambda(r)}} \Bigg[ 2r f_{QQ} Q' \left( e^{\lambda(r)} -1 \right) + f_Q \Big[ \left( e^{\lambda(r)} -1 \right) \left( 2 + r \nu'(r) \right) \\
        + \left( e^{\lambda(r)} + 1 \right)r \lambda'(r)  \Big] + f\ r^2\ e^{\lambda(r)} \Bigg],
    \end{multline}
    \begin{multline}
        \label{pr1}
        8\pi  P_r(r) = - \frac{1}{2r^2\ e^{\lambda(r)}} \Bigg[ 2r f_{QQ} Q' \left( e^{\lambda(r)} -1 \right) + f_Q \Big[ \left( e^{\lambda(r)} -1 \right) ( 2 + r \nu'(r) \\
        + r \lambda'(r) ) + 2r \nu'(r)  \Big] + f\ r^2\ e^{\lambda(r)} \Bigg],
    \end{multline}
    \begin{multline}
        \label{pt1}
        8\pi  P_{\perp}(r) = - \frac{1}{4r\ e^{\lambda(r)}} \Bigg[- 2r f_{QQ} Q' \nu'(r) + f_Q \Big[ 2\nu'(r) \left( e^{\lambda(r)} -1 \right) - r \nu'(r)^2 \\
        + \lambda'(r) \left( 2e^{\lambda(r)} + r\nu'(r) \right) - 2r \nu''(r)  \Big] + 2f\ r^2\ e^{\lambda(r)} \Bigg],
    \end{multline}
where $f_{QQ} \equiv \frac{df_Q}{dQ}$ and prime ($'$) denotes derivative with respect to $r$. The analytical expression for the non-metricity tensor can be obtained using equation \ref{nonmetricity} and equation \ref{lineelement} as 
    \begin{equation}
        \label{nonmetricity1}
        Q = \frac{1}{r} \left( \nu'(r) + \lambda'(r) \right) \left( e^{-\lambda(r)} -1 \right).
    \end{equation}
We consider the following linear form of $f(Q)$ gravity
    \begin{equation}
        \label{fQ}
        f(Q) = \alpha \ Q + \phi,
    \end{equation}
where $\alpha$ is an arbitrary non-zero  constant and $\phi$ is the  cosmological constant, i.e., $\phi = 2.036 \ \times \ 10^{-35} s^{-2}$ \cite{Carmeli2001ValueExperiment}. For $\alpha = 0 = \phi$, Eq. (\ref{fQ}) reduces to, $f(Q) = Q$, which is the Symmetric Teleparallel Equivalent of General Relativity (STEGR) case.  % Different phases of the universe is based on different values of this constant, i.e., a positive value means an accelerated universe, while a negative value indicates a slowing down of the expansion of the universe. In this work we will consider a negative value of the cosmological parameter following the new findings by DESI [CITE : arxiv:2503.14745v1, arxiv:2503.14739v2, arxiv:2503.14738v2].  
Now, using the analytical expression of the non-metricity tensor and the linear form of $f(Q)$ gravity, the field equations can be rewritten as
    \begin{equation}
        \label{rho2}
        8\pi \rho(r) = \frac{2\alpha + \phi \ r^2 + 2\alpha \ e^{\lambda(r)} \Big( \lambda'(r) - 1  \Big)}{2 \ r^2},
    \end{equation}
    \begin{equation}
        \label{pr2}
        8\pi  P_r(r) = \frac{ - 2\alpha + \phi \ r^2 - 2\alpha \ e^{\lambda(r)} \Big( \nu'(r) + 1  \Big)}{2 \ r^2},
    \end{equation}
    \begin{equation}
        \label{pt2}
        8\pi  P_{\perp}(r) = \frac{\alpha \ e^{\lambda(r)}}{2} \Bigg[ \nu''(r) + \frac{(\nu')^2}{2} - \frac{\nu'(r) \ \lambda'(r)}{2} +\frac{(\nu'(r) - \lambda'(r))}{r} \Bigg]  - \frac{\phi}{2} .
    \end{equation}

%%%%%%%%%%%%%%%%%%%%%%%%%%%%%%%%%%%%%%%%%%%%%%%%%%%%%%%%%%%%%%%%%%%%

\section{Exact solution of field equations in pseudo-spherical geometry}
\label{sec:4}
In order to solve the field equations, we take a physically stable form of the metric potential ($\lambda(r)$) given by the pseudo-spheroidal geometry \cite{Tikekar1998RelativisticSpace-time}:
    \begin{equation}
        \label{metric}
        e^{\lambda(r)} = \frac{1 + \mu \ \frac{r^2}{L^2}}{1 + \frac{r^2}{L^2}},
    \end{equation}
Here $\mu$ is the spheroidicity parameter, and $L$ is a geometrical parameter. Now, using the Durgapal-Banerjee transformation \cite{Durgapal1983NewRelativity} on the metric potentials, we introduce new functions $y(x)$ and $z(x)$ with an independent variable $x$:
    \begin{equation}
        \label{db}
        x = r^2 / L^2, \;\;\;\;\; z(x) = e^{-\lambda(r)}, \;\;\;\;\; y^2(x) = e^{\nu(r)}.
    \end{equation}
Following this transformation, the field equations given in Eqs. (\ref{rho2}) - (\ref{pt2}) takes the form as
     \begin{equation}
        \label{density}
        8\pi \ \rho(x) = \frac{2\alpha + L^2 \phi \ x - 2\alpha  \Big( z(x) + 2x\ z'(x) \Big)}{2  x \ L^2 },
    \end{equation}
    \begin{equation}
        \label{radialpressure}
        8\pi  P_r(x) = \frac{ y(x) \Big( 2\alpha \left( z(x) - 1 \right) - L^2 \phi \ x \Big) + 8\alpha x \ z(x) \ y'(x)}{2x \ L^2 \ y(x)},
    \end{equation}
    \begin{equation}
        \label{tangentialpressure}
        8\pi  P_{\perp}(x) = \frac{ 8\alpha \ z(x) \Big( y'(x) + x \ y''(x) \Big) + 4\alpha x \ z'(x) \ y'(x) + y(x) \Big( 2\alpha \ z'(x) - L^2 \phi \Big) }{2  L^2 \ y(x)}.
    \end{equation}
Subsequently the pseudo-spheroidal geometry will take the form
    \begin{equation}
        \label{metric1}
        z(x) = \frac{1+x}{1 + \mu \ x}.
    \end{equation}
To close the system, we need to choose one additional constraint, which is given by the relation between the radial pressure ($P_r$) and the energy density ($\rho$), i.e., the equation of state (EoS). We consider a linear EoS given by the MIT Bag model for describing the strange quark matter (SQM) distribution in an anisotropic star as:
    \begin{equation}
        \label{eos}
        P_r = \frac{1}{3} (\rho - 4 \mathcal{B}_g),
    \end{equation}
where, $\mathcal{B}_g$ is the Bag constant of units MeV/fm$^3$. 

Using the above set of equations given in Eqs. (\ref{density}) - (\ref{eos}), we obtain the expression for another metric potential, i.e.,
    \begin{multline}
        \label{metric2}
        y(x) = D \exp \Bigg[ \frac{1}{6 \alpha}  \Bigg( \log (x+1) \Big(\alpha  (2 \mu -3) + 16 \pi \ \mathcal{B}_g (\mu -1) L^2-(\mu -1) L^2 \phi \Big) \\
        + \mu \ L^2 x (\phi -16 \pi \ \mathcal{B}_g ) +\alpha  \log (\mu \ x+1) \Bigg) \Bigg]
    \end{multline}
Using this in Eqs. (\ref{density}) - (\ref{tangentialpressure}), we obtain the analytical expression for the energy density ($\rho$), the radial pressure ($P_r$), and the transverse pressure ($P_{\perp}$) taking the form
    \begin{equation}
        \label{densityf}
        \rho(x) = \frac{1}{16\pi} \Bigg[ \frac{2 \alpha  (\mu -1) (\mu  x+3)}{(\mu  L x+L)^2}+\phi \Bigg],
    \end{equation}
    \begin{equation}
        \label{radpref}
        P_r(x) = \frac{1}{3} \Big( \rho - 4 \mathcal{B}_g \Big),
    \end{equation}
    \begin{multline}
        \label{tranpref}
        P_{\perp}(x) = \frac{1}{144 \pi \ \alpha  L^2 (x+1) (\mu \ x+1)^3} \Bigg[ 512 \pi^2 x \  \mathcal{B}_g^2 (\mu \ L x+L)^4 - 32 \pi \ \mathcal{B}_g (\mu \ L x+L)^2  \\
        \Big( 2 x \ \phi  (\mu \ L x+L)^2+\alpha  (x (\mu  ((4 \mu +2) x+15)-3)+6) \Big) + \alpha \ \phi  \Big(x (\mu  ((8 \mu -5) x+21)-15)+3 \Big) \\
        \times \ (\mu \ L x+L)^2+2 x \ \phi ^2 (\mu \ L x+L)^4+2 \alpha ^2 (\mu -1) \Big(x (\mu  (x (2 \mu  (2 (\mu -1) x+9)-21)+15)-9)+9 \Big) \Bigg]
    \end{multline}

%%%%%%%%%%%%%%%%%%%%%%%%%%%%%%%%%%%%%%%%%%%%%%%%%%%%%%%%%%%%%%%%%%%%

\section{Exterior metric and boundary conditions}
\label{sec:5}
To obtain the exterior vacuum solution in the case of $f(Q)$ gravity, we need to study the spherical vacuum first with $\mathcal{T}_{\eta \kappa} = 0$. In vacuum case, a linear combination of the field equations given in Eqs. (\ref{rho1}) - (\ref{pt1}) yields the following:
    \begin{align}
        \begin{aligned}
             &\frac{e^{-\lambda(r)}f_{Q} \Big( \nu'(r) + \lambda'(r) \Big)}{r} = 0\\
            \Rightarrow\,\,\, & \nu'(r) + \lambda'(r)  = 0
        \end{aligned}
    \end{align}
This leads to $e^{\nu(r)} = e^{-\lambda(r)}$, where the integration constant can be ignored by rescaling the time coordinate. As a result, the non-metricity scalar, given in Eq. (\ref{nonmetricity1}), vanished in the context of a spherical vacuum. Now, evaluating the field equations given in Eqs. (\ref{rho1}) and (\ref{pt1}) under vacuum condition, we obtain
    \begin{equation}
        2f_{Q0} \left( e^{\lambda(r)} - 1 + r\, \lambda'(r) \right) + f_{0} r^2 e^{\lambda(r)} = 0,
    \end{equation}
where $f_{0} = f(Q)|_{Q=0}$ and $f_{Q0} = f_{Q}|_{Q=0}$ are constants. Solving this equation we get the exterior metric as
    \begin{equation}
        \label{ext1}
        e^{\nu(r)} = e^{-\lambda(r)} = 1 + \frac{C_{0}}{r} + \frac{f_{0}}{6f_{Q0}} r^2,
    \end{equation}
where $C_{0}$ is an integration constant. This is the Schwarzschild-de Sitter (SdS) solution with the cosmological constant $\Lambda = \frac{f_{0}}{2f_{Q0}}$. For a linear form of $f(Q)$, as given in Eq. (\ref{fQ}), this cosmological constant term can be written as, $\Lambda = \frac{\phi}{2 \alpha}$. Now,  putting $C_{0} = -2M$ in Eq. (\ref{ext1}), the exterior metric reduces to the Schwarzschild-de Sitter (SdS) solution:
    \begin{equation}
        e^{\nu(r)} = e^{-\lambda(r)} = 1 - \frac{2M}{r} + \frac{\Lambda}{3} r^2.
    \end{equation}
Therefore, for a static uncharged compact star with mass $M$, the exterior vacuum solution for a linear form of $f(Q)$ is given by the Schwarzschild-de Sitter (SdS) line element as
    \begin{equation}
        \label{exmetric}
        ds^2 = -\Big( 1- \frac{2M}{r}  + \frac{\Lambda}{3} r^2 \Big)\, dt^2 + \Big( 1- \frac{2M}{r}  + \frac{\Lambda}{3} r^2 \Big)^{-1}\,dr^2 + r^2 \left( d\theta^2 + \sin^2\theta d\phi^2 \right).
    \end{equation}
So, the continuity of the metric potentials over the boundary ($r=R$) of the strange quark star leads to \cite{Darmois1927,Israel1967,Bonnor1981},
    \begin{equation}
        \label{bc1}
        g_{rr} = \Big( 1- \frac{2M}{R}  + \frac{\Lambda}{3} R^2 \Big)^{-1} = e^{\lambda(R)},
    \end{equation}
    \begin{equation}
        \label{bc2}
        g_{tt} =\Big( 1- \frac{2M}{R}  + \frac{\Lambda}{3} R^2 \Big) = e^{\nu(R)}. 
    \end{equation}
In addition to this, the radial pressure must vanish at the surface ($r=R$) of the strange quark star, i.e., $P_r(R) = 0$ \cite{Bonnor1981}, which gives us another boundary condition:
    \begin{equation}
        \label{bc3}
        \frac{1}{48\pi} \Bigg[ \frac{2 \alpha  (\mu -1) (\mu  x+3)}{(\mu  L x+L)^2}+\phi \Bigg] - \frac{4}{3} \mathcal{B}_g = 0.
    \end{equation}
By solving these boundary conditions simultaneously, we obtain the expressions for the model parameters $L$, $D$, and $\mathcal{B}_g$ as follows,
    \begin{equation}
    \label{bc4}
        L = \Bigg[ \frac{R^2 \Big(3 (\mu -1) R - \mu \left( 6  M - \Lambda   R^3\right) \Big)}{6 M-\Lambda  R^3} \Bigg]^{\frac{1}{2}},
    \end{equation}
    \begin{equation}
        D = \frac{e^{-\frac{R^2 \mu  (-16 \text{Bg} \pi +\phi )}{6 \alpha }} \left(1+\frac{R^2}{L^2}\right)^{\frac{\alpha  (3-2 \mu )-16 \text{Bg} L^2 \pi 
   (-1+\mu )+L^2 (-1+\mu ) \phi }{6 \alpha }} \left(3 R-6 M+R^3 \Lambda \right)}{3 R \left(1+\frac{R^2 \mu }{L^2}\right)^{1/6}}
    \end{equation}
    \begin{equation}
      \label{bc5}
        \mathcal{B}_g = \frac{1}{576 \pi} \Bigg[\frac{2 \alpha  \left(6 M - \Lambda  R^3\right) \Big(9R (\mu -1)  - 2\mu \left(6  M - \Lambda   R^3\right) \Big)}{ R^4(\mu -1)} + 9\phi \Bigg],
    \end{equation}
%    \begin{equation}
%        D =  \frac{\sqrt{(R-2M)}}{\Bigg[ R \exp \Bigg( \frac{1}{6 \alpha}  \Big( \mu  R^2 (\phi -16 \pi \ \mathcal{B}_g )+\alpha  \log \left(\frac{\mu \ R^2}{L^2}+1\right) + \zeta \log \left(\frac{R^2}{L^2}+1\right) \Big) \Bigg) \Bigg]^{\frac{1}{2}} },
%    \end{equation}
%where $\zeta = \Big( \alpha  (2 \mu -3)+16 \pi \ \mathcal{B}_g (\mu -1) L^2-(\mu -1) L^2 \phi \Big)$. 
Therefore, for the linear $f(Q)$, with fixed value of the $\alpha$ and $\phi$, we must assume one {\it ad hoc} relation to construct the stellar model. We choose different values of the spheroidicity parameter ($\mu$) to explore the physical features of the stellar model in linear $f(Q)$ gravity.. 

%%%%%%%%%%%%%%%%%%%%%%%%%%%%%%%%%%%%%%%%%%%%%%%%%%%%%%%%%%%%%%%%%%%%

\section{Criteria for a Realistic and Stable Stellar Model}
\label{sec:6}

A physically realistic and stable stellar configuration must satisfy the following essential conditions \cite{Delgaty1998}:

\begin{enumerate}
    \item Regularity and Monotonicity of Physical Quantities: 
    The energy density $\rho$ and pressure $P$ must be positive and finite throughout the star. At the center ($r = 0$), they should take finite central values:
    \[
    \rho(0) = \rho_c \quad \text{and} \quad P(0) = P_c
    \]
    Additionally, both $\rho$ and $P$ must decrease monotonically with radial distance:
    \[
    \frac{d\rho}{dr} < 0 \quad \text{and} \quad \frac{dP}{dr} < 0
    \]

    \item Isotropy at the Center: 
    At the center of the star, radial pressure $P_r$ and tangential pressure $P_\perp$ must be equal, ensuring isotropy:
    \[
    P_r(0) = P_\perp(0)
    \]

    \item Boundary Condition:
    The surface of the star is defined by the radius $R$ where the radial pressure vanishes:
    \[
    P_r(R) = 0
    \]

    \item Causality Condition:
    The speed of sound inside the star must be less than the speed of light in both radial and tangential directions. Hence,
    \[
    0 \leq v_r^2 = \frac{dP_r}{d\rho} < 1 \quad \text{and} \quad 0 \leq v_\perp^2 = \frac{dP_\perp}{d\rho} < 1
    \]

    \item Adiabatic Index and Stability:
    The adiabatic index $\Gamma$ must exceed a critical value $\Gamma_{\text{crit}}$ throughout the interior of the star to maintain dynamical stability:
    \[
    \Gamma = \left(1 + \frac{\rho}{P} \right) \frac{dP}{d\rho} > \Gamma_{\text{crit}}
    \]

    \item Energy Conditions:
    The energy-momentum tensor of the matter distribution must satisfy the following standard energy conditions: Null Energy Condition (NEC), Weak Energy Condition (WEC), Strong Energy Condition (SEC), Dominant Energy Condition (DEC), and Trace Energy Condition (TEC).
    
\end{enumerate}

%%%%%%%%%%%%%%%%%%%%%%%%%%%%%%%%%%%%%%%%%%%%%%%%%%%%%%%%%%%%%%%%%%%%

\section{Physical Analysis}
\label{sec:7}
In this section, we investigate the physical viability of the stellar model through analytical and graphical analysis for a given mass ($M$) and radius ($R$) of a compact object. We consider the compact object EXO 1745-248 \cite{Ozel2009THE1745248} with observed mass $M = 1.7\, M_{\odot}$ and predicted radius $R = 9\, km$. According to the Ref. \cite{Chattopadhyay2010RelativisticSpace-time}, stable realistic solutions are possible for an uncharged star in four dimensions for (i) $\mu < - \frac{3}{17}$ and (ii) $\mu > 5$. In the present paper, we stick to the second case only, i.e., $\mu > 5$. 

%%%%%%%%%%%%%%%%%%%%%%%%%%%%%%%%%%%%%%%%%%%%%%%%%%%%%%%%%%%%%%%%%%%%

\subsection{Constrain on $\alpha$}
\begin{figure}[tbp]
        \centering
        \includegraphics[width=0.5\linewidth]{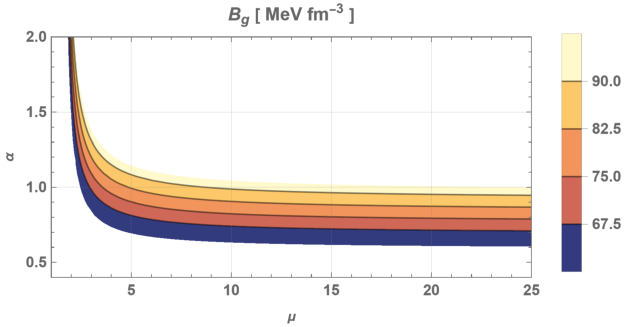}
        \caption{Contour of $B_g$ within stable limit for different values of $\alpha$ and $\mu$ for the compact object EXO 1745-248 ($M = 1.7\, M_{\odot}$ and $R = 9\, km$).}
        \label{fig:alphaconstraint}
    \end{figure}

In this subsection, we constrain the values of $\alpha$ using a viable physical limit on the Bag constant $B_g$. For the stability of strange quark matter, the Bag constant is bound by a certain limit. The lower bound on $B_g$ is 145 $MeV$, which is equivalent to 57.55 $MeV\, fm^{-3}$ \cite{Madsen1998PhysicsMatter}. On the other hand, the upper bound on $B_g$ is 162.8 $MeV$ or, equivalently, 95.11 $MeV\, fm^{-3}$ \cite{Madsen1998PhysicsMatter}. Using Eq. (\ref{bc5}), we can constraint the values of $\alpha$ for a certain compact object with mass $M$ and radius $R$ (EXO 1745-248 in our case). The contour plot of $B_g$ in Fig. (\ref{fig:alphaconstraint}) shows the range $\in [57.55,95.11]\,MeV\,fm^{-3}$ for different values of $\alpha$ and $\mu$. We note that for a range of $\alpha$ for given $\mu$, the Bag constant is within the stable limit. In the present paper, we only choose $\alpha = 0.85$ to explore the physical features of the stellar model for $\mu > 5$.

%%%%%%%%%%%%%%%%%%%%%%%%%%%%%%%%%%%%%%%%%%%%%%%%%%%%%%%%%%%%%%%%%%%%

\subsection{Constrain on Mass-Radius Relation}
\begin{figure}[tbp]
        \centering
        \includegraphics[width=0.6 \linewidth]{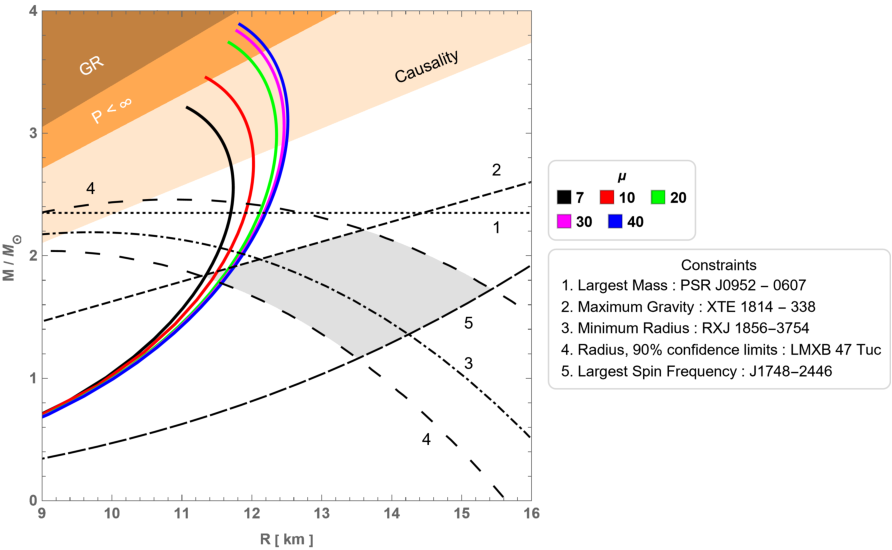}
        \caption{Mass-Radius profile for surface density ($\rho_s$) = 4.7 $\times$ $10^{14}\ gm\, cm^{-3}$. }
        \label{fig:MRconstraints}
    \end{figure}

We plot the theoretical Mass-Radius (M-R) relationship in Fig. (\ref{fig:MRconstraints}), with the observable constraints represented by dotted and dashed lines. The latest stringent constraints on the mass-radius relation of neutron stars are determined by the maximum mass, minimum radius, highest rotational frequency, and maximum surface gravity observed in pulsars \citep{Trumper2011ObservationsDensities}. The mass of the 3.15 millisecond pulsar PSR J0952-0607, $M =$ 2.35 $\pm$ 0.17 $M_{\odot}$, represents the highest neutron star mass determined with significant experimental confidence, derived from the Keplerian orbital parameters and Shapiro time delay of the binary system \citep{Romani2022PSRStar}. The thermonuclear burst oscillation light curves from the accreting millisecond pulsar XTE J1814-338 are analyzed to establish an upper limit on the neutron star's surface gravity, thereby constraining the mass-radius relation, $\frac{M}{R} < 0.24$ \citep{Bhattacharyya2005ConstraintsJ1814338}. Through the analysis and fitting of the optical and X-ray spectra of the source, a lower bound for the radius of RX J1856-3754, as perceived by an observer at infinity, has been established: $R_{\infty} = R (1-\frac{2M}{R})^{-1/2} > 16.8$ km. This yields the constraint $\frac{2M}{R} > (1-\frac{R^2}{R_{\infty}^2})$, with $R_{\infty} = 16.8$. \citep{Trumper2004TheStar}. The Chandra data for the low-mass X-ray binary X7 is analyzed under the assumption of a neutron star with a mass of $M=1.4 M_{\odot}$, which yields a neutron star radius of $R = 14.5^{+1.8}_{-1.6}$ km at a $90\%$ confidence level. \citep{Heinke2006ATucanae}. Considering both general relativity and deformations, the maximum spin frequency of a neutron star is determined to be $\nu_{max} = 1045(\frac{M}{M_{\odot}})^{1/2}(\frac{10 \;km}{R})^{3/2}$ Hz, which remains unaffected by the equation of state. \citep{Lattimer2004TheStars}. The fastest recorded pulsar is PSR J1748-2246ad with a rotation frequency of 716 Hz \citep{Hessels2006AHz}, which imposes a constraint on the M-R relation as, $M \geq 0.47 (\frac{R}{10 \;km})^3 \; M_{\odot}$. The Brown region, labeled by ``GR'', is excluded by the constraint $R > \frac{2GM}{c^2}$, the orange region, labeled by ``$P < \infty$'', is excluded by the finite pressure constraint $R > \frac{9}{4}\frac{GM}{c^2}$, and the light orange region, labeled by ``causality'', is excluded by the constraint ,$R > \frac{2.9GM}{c^2}$ \cite{LATTIMER2007NeutronConstraints}. The allowed M-R relations should pass through the area delimited by the constraints marked from line no. 2-5 (gray region) in the Fig. (\ref{fig:MRconstraints}) and it must have a maximum mass larger than the mass of PSR J0952-0607, $M =$ 2.35 $\pm$ 0.17 $M_{\odot}$ marked by the dotted line (line 1). It is found that for $\mu \geq 7$ only, we get allowed M-R relation, that is passing through the area delimited by the constraints. 
%It is also clear from the figure that with increasing value of the spheroidal parameter $\mu$, the maximum mass in increasing.
    
%    \begin{table}
%        \centering
%        \begin{tabular}{cccccllccccc} 
%             Compact &  Measured &  Observed &  Measured &   \multicolumn{8}{c}{Predicted Radius and compactness from the model}\\
%             Objects&  Mass &  Radius &  Compactness &   \multicolumn{2}{c}{$\mu$ = 10}&\multicolumn{2}{c}{$\mu$ = 20}&  \multicolumn{2}{c}{$\mu$ = 30}&  \multicolumn{2}{c}{$\mu$ = 40}\\
% & ($M_{\odot}$)& ($R$)& ($c$)& $R$ (km)& $c$& $R$ (km)& $c$& $R$ (km)& $c$& $R$ (km)&$c$\\
%             4 U 1608 -  52&  1.74&  9.3&  0.275968&   11.34& 0.2263&11.47&  0.223758&  11.495&  0.223271&  11.52& 0.222786\\
%             PSR J1903 + 0327&  1.667&  9.438&  0.260524&   & &&  &  &  &  & \\
%             PSR J1614 - 2203&  1.97&  11&  0.264159&   & &&  &  &  &  & \\
%             4 U 1820 -  30&  1.58&  9.1&  0.256099&   & &&  &  &  &  & \\
%             &  &  &  &   & &&  &  &  &  & \\ 
%        \end{tabular}
%        \caption{Caption}
%        \label{tab:MR_data}
%    \end{table}

%%%%%%%%%%%%%%%%%%%%%%%%%%%%%%%%%%%%%%%%%%%%%%%%%%%%%%%%%%%%%%%%%%%%

\subsection{Regularity of the metric potentials}
\begin{figure}[tbp]
        \centering
        \includegraphics[width=0.48\linewidth]{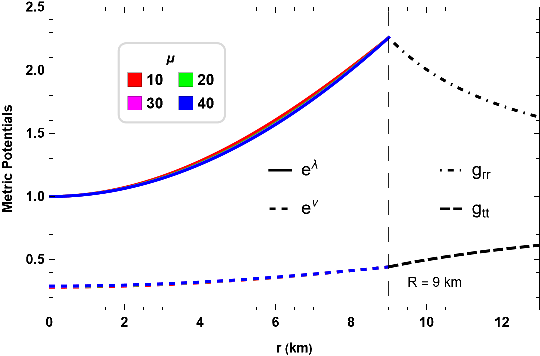}
        \caption{Radial profile of the metric potentials for the compact object EXO 1745-248 ($M = 1.7\, M_{\odot}$ and $R = 9\, km$)}
        \label{fig:metrics}
    \end{figure}

The radial profile of the metric potentials ($e^{\nu(r)}$ and $e^{\lambda(r)}$) are plotted in Fig. (\ref{fig:metrics}) for the compact star EXO 1745-248. It is evident that the metric potentials are finite at the center of the star, i.e., $e^{\nu(0)} = e^{\lambda(0)} = $ const.. The metric potentials are matching with the Schwarzschild-de Sitter (SdS) at the boundary of the star ($r = R$). It is also noted that $(e^{\nu(r)})'_{r=0} = (e^{\lambda(r)})'_{r=0} = 0$, which indicates that the metric potentials are regular inside the stellar configuration \cite{Delgaty1998,Leibovitz1969} . 

%%%%%%%%%%%%%%%%%%%%%%%%%%%%%%%%%%%%%%%%%%%%%%%%%%%%%%%%%%%%%%%%%%%%

\subsection{Energy density and pressure}

    \begin{figure}[tbp]
        \centering
        \includegraphics[width=0.48\linewidth]{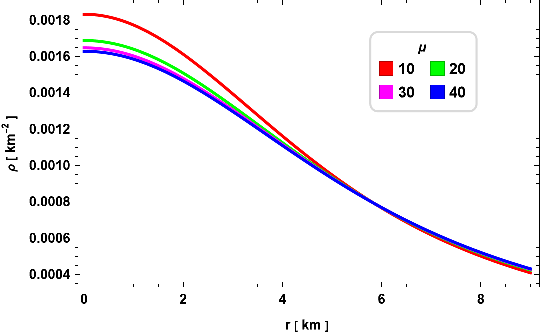}\,\,\,\,
        \includegraphics[width=0.48\linewidth]{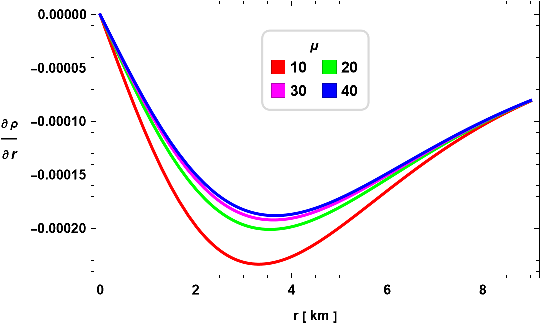}
        \caption{Left: Radial variation of the energy density ($\rho$) and Right: radial variation of the gradient of the energy density ($\frac{\partial\rho}{\partial r}$) for the compact object EXO 1745-248 ($M = 1.7\, M_{\odot}$ and $R = 9\, km$)}
        \label{fig:density}
    \end{figure}
    
\begin{figure}[h]
        \centering
        \includegraphics[width=0.48\linewidth]{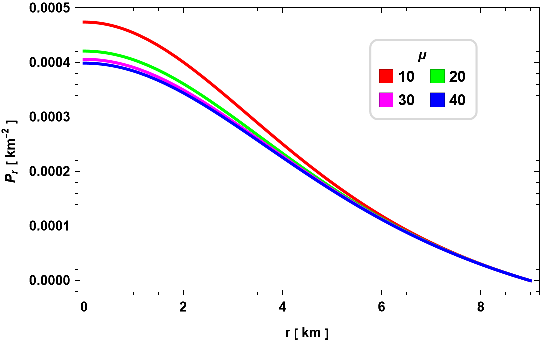} \,\,\,\,
        \includegraphics[width=0.48\linewidth]{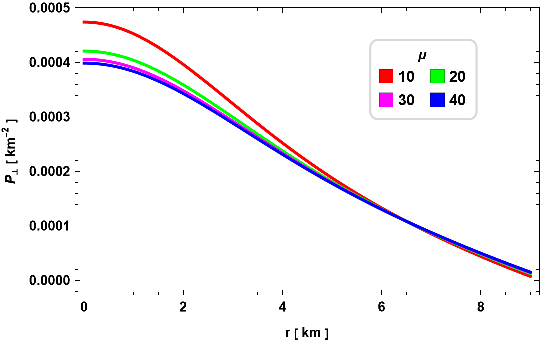}
        \caption{Left: Radial variation of the radial pressure ($P_r$) and Right: radial variation of tangential pressure ($P_{\perp}$) for the compact object EXO 1745-248 ($M = 1.7\, M_{\odot}$ and $R = 9\, km$)}
        \label{fig:pressure}
    \end{figure}

In this subsection, we have obtained the central values of the energy density and pressure profiles, and analyzed the radial variation of energy density, pressure profiles and their gradient throughout the star. The analytical expression for the central energy density and the central pressure is obtained as,
    \begin{equation}
        \label{centraldensity}
        \rho_c =  \frac{6 \alpha  (\mu-1 )}{16 \pi L^2} + \frac{\phi}{16 \pi } > 0.
    \end{equation}
    \begin{equation}
        \label{centralpressure}
        P_c =  \frac{1}{48 \pi } \Bigg(   \frac{6 \alpha  (\mu-1 )}{L^2}+\phi -64 \pi  B_{g} \Bigg) > 0.
    \end{equation}
The radial variation of the energy density ($\rho$) and the gradient of energy density ($\frac{\partial\rho}{\partial r}$) is plotted in Fig. (\ref{fig:density}) for the compact star EXO 1745-248 for different values of spheroidicity parameter ($\mu$). It is evident that the energy density is finite through out the star. It is maximum at the center and monotonically decreasing towards the boundary ($r=R$) of the star. It is noted that radial variation of the gradient of energy density ($\frac{\partial\rho}{\partial r}$) is less than zero for all the values of $\mu$ considered. Fig. (\ref{fig:pressure}) shows the radial profile of the radial pressure  ($P_r$) and the tangential pressure ($P_{\perp}$) inside the star for different values of $\mu$ considered. It is clear form the figure that both the pressures are equal and maximum at the center of the star. We note that both the pressure profiles are monotonically decreasing away from the center for all the values of $\mu$. The radial pressure vanished at the boundary of the star ($r=R$), whereas the tangential pressure remains non-zero, giving rise to anisotropy. The radial variation of the gradient of both the pressure profiles are less than zero throughout the stellar interior. It is also important to note that for increasing value of the spheroidicity parameter $\mu$, the energy density and pressure decreases at the center of the star. However, the gap between the central values of energy density and pressure for different values of $\mu$ is decreasing with an increase in $\mu$. Central and surface values of the energy density and pressure for different values of spheroidicity parameter ($\mu$) are tabulated in Table \ref{tab:physicalparameters}.

    \begin{figure}[t]
        \centering
        \includegraphics[width=0.48\linewidth]{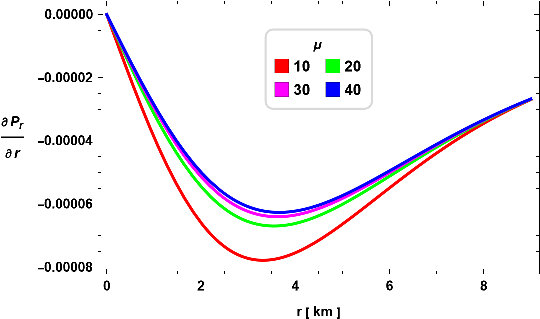}\,\,\,\,
        \includegraphics[width=0.48\linewidth]{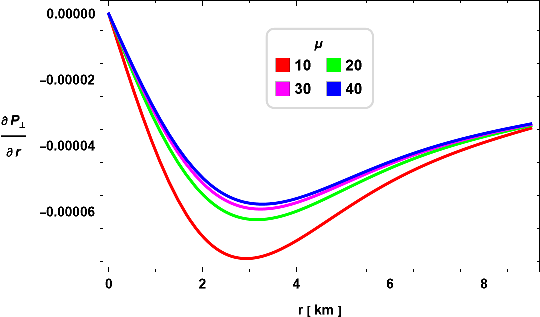}
        \caption{Left: Radial variation of the gradient of radial pressure ($P_r$) and Right: radial variation of the gradient of tangential pressure ($P_{\perp}$) for the compact object EXO 1745-248 ($M = 1.7\, M_{\odot}$ and $R = 9\, km$)}
        \label{fig:gradpressure}
    \end{figure}

     \begin{table}[h]
        \centering
        \begin{tabular}{c c c c c c c}
           \hline
           $\mu$    &    $L$   &  $D$   &   $B_{g}$   &   $\rho_c$    &   $\rho_R$    &   $P_c$  \\
                &   (km$^{2}$) &        & (MeV fm$^{-3}$) & (gm cm$^{-3}$) & (gm cm$^{-3}$) & (gm cm$^{-3}$) \\ \hline
            10  & 22.3221 & 0.281822 & 77.4536 & 2.4678 $\times \, 10^{15}$ & 5.5195 $\times \, 10^{14}$ & 6.3862 $\times \, 10^{14}$ \\ 
            20  & 33.7922 & 0.287922 & 80.3189 & 2.2733 $\times \, 10^{15}$ & 5.7237 $\times \, 10^{14}$ & 5.6698 $\times \, 10^{14}$ \\ 
            30  & 42.2558 & 0.289734 & 81.2081 & 2.2190 $\times \, 10^{15}$ & 5.7871 $\times \, 10^{14}$ & 5.4678 $\times \, 10^{14}$ \\ 
            40  & 49.2868 & 0.290604 & 81.6413 & 2.1935 $\times \, 10^{15}$ & 5.8179 $\times \, 10^{14}$ & 5.3724 $\times \, 10^{14}$  \\ \hline
        \end{tabular}
        \caption{Tabulation of model parameters and some physical quantities at the center and surface for $\alpha = 0.85$, $\phi = 2.036 \times 10^{-35}$, and different values of spheroidicity parameter ($\mu$) for compact object EXO 1745-248 ($M = 1.7\, M_{\odot}$ and $R = 9\, km$)}
        \label{tab:physicalparameters}
    \end{table}

%%%%%%%%%%%%%%%%%%%%%%%%%%%%%%%%%%%%%%%%%%%%%%%%%%%%%%%%%%%%%%%%%%%%

\subsection{Anisotropy}
\begin{figure}[tbp]
        \centering
        \includegraphics[width=0.5\linewidth]{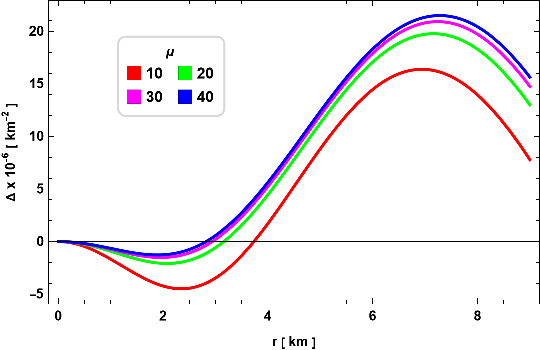}
        \caption{Radial variation of the anisotropy factor ($\Delta$) for the compact object EXO 1745-248 ($M = 1.7\, M_{\odot}$ and $R = 9\, km$)}
        \label{fig:anisotropy}
    \end{figure}

The anisotropy factor ($\Delta$) is defined as follows,
    \begin{equation}
        \label{anisotropy}
        \Delta = P_{\perp} - P_r.
    \end{equation}
The anisotropy inside the star refers to the difference between the radial pressure ($P_r$) and tangential pressure ($P_{\perp}$). This pressure anisotropy produces an additional force gradient in the star's hydrostatic equilibrium. $P_{\perp} > P_r$, i.e. a positive anisotropy denotes a outward pressure resulting in an increase in the repulsive force inside the star. Whereas, $P_{\perp} > P_r$ signifies the increase in the inward attractive force in the hydrostatic equilibrium \cite{Gokhroo1994}. Fig. (\ref{fig:anisotropy}) shows the the radial variation of the anisotropy factor for different values of the spheroidicity parameter $\mu$. It is noted that the stellar model maintains isotropy at the center of the star for any given value of $\mu$. As we move away from the center of the star, $\Delta$ is negative in nature up to a certain radial distance, where anisotropy vanishes. After that $\Delta$ starts increasing again and reached a maximum near the boundary of the star. This feature of the anisotropy factor is the same for all values of $\mu$ considered.

%%%%%%%%%%%%%%%%%%%%%%%%%%%%%%%%%%%%%%%%%%%%%%%%%%%%%%%%%%%%%%%%%%%%

\subsection{Energy conditions}
\begin{figure}[tbp]
    \centering
    \begin{tabular}{c c}
    \includegraphics[width=0.5\linewidth]{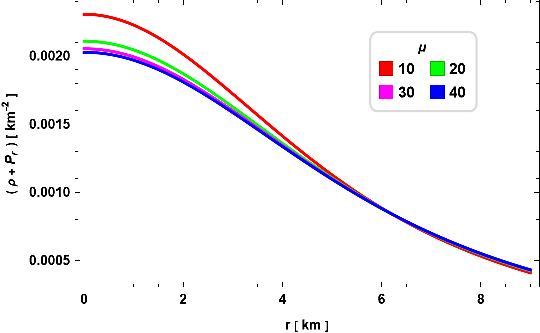} &
    \includegraphics[width=0.5\linewidth]{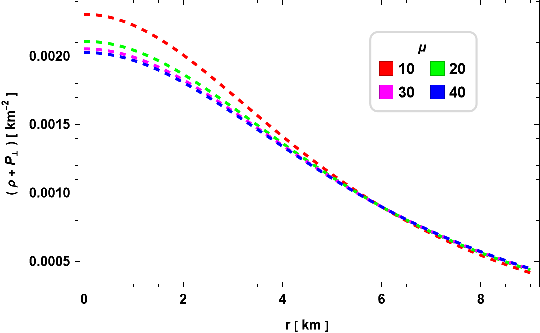}  \\
    \includegraphics[width=0.5\linewidth]{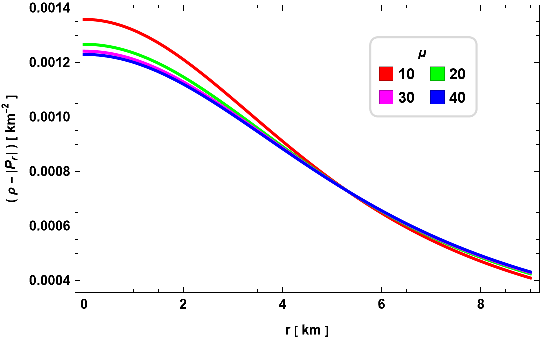} &
    \includegraphics[width=0.5\linewidth]{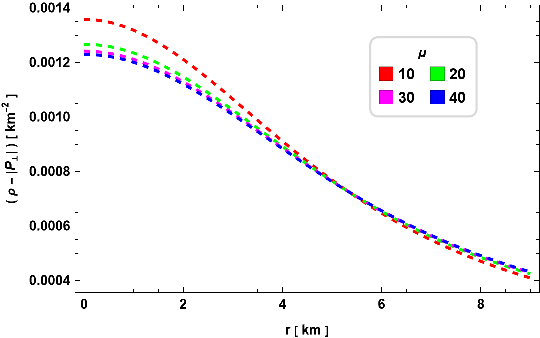} \\
    \includegraphics[width=0.5\linewidth]{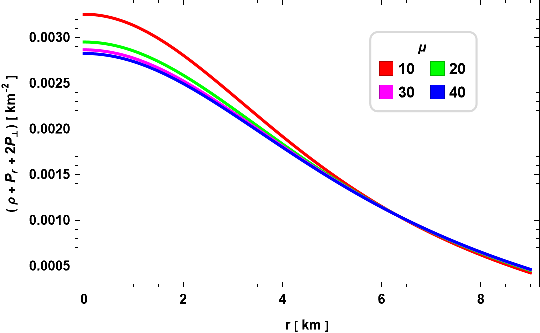} &
    \includegraphics[width=0.5\linewidth]{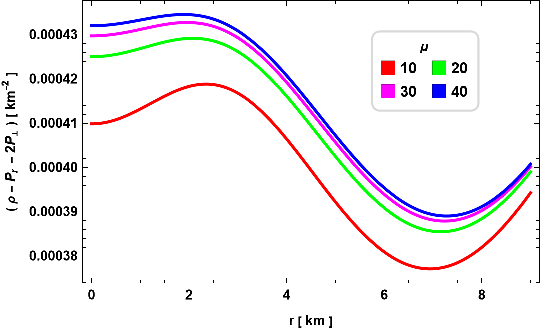}
    \end{tabular}
    \caption{Radial variation of the energy conditions for the compact object EXO 1745-248 ($M = 1.7\, M_{\odot}$ and $R = 9\, km$)}
    \label{fig:energyconditions}
\end{figure}

Here we have analyzed the following energy conditions \cite{Witten1981,Visser1997,Mandal2020}:
    \begin{itemize}
        \item Null Energy Condition (NEC): $\rho + P_i \geq 0$
        \item Weak Energy Condition (WEC): $\rho \geq 0$, \quad $\rho + P_i \geq 0$
        \item Dominant Energy Condition (DEC): $\rho \geq |P_i|$
        \item Strong Energy Condition (SEC): $\rho + P_i \geq 0$, \quad $\rho + \sum_i P_i \geq 0$
        \item Trace Energy Condition (TEC): $\rho - \sum_i P_i \geq 0$
    \end{itemize}
where $P_i$ denotes either radial or tangential pressure. Fig. (\ref{fig:energyconditions}) shows the radial variation of the energy conditions inside the stellar interior. It is evident from the figure that all the energy conditions are satisfied for all the values of the spheroidicity parameter considered.
    
%%%%%%%%%%%%%%%%%%%%%%%%%%%%%%%%%%%%%%%%%%%%%%%%%%%%%%%%%%%%%%%%%%%%

\section{Stability Analysis}
\label{sec:8}
The stability of the stellar model is assessed through the following criteria: (1) Stability analysis on sound speed, (2) Adiabatic index, (3) Zel'dovich condition, and (4) Hydrostatic equilibrium.
%%%%%%%%%%%%%%%%%%%%%%%%%%%%%%%%%%%%%%%%%%%%%%%%%%%%%%%%%%%%%%%%%%%%

\subsection{Stability analysis on sound speed}
\begin{figure}[tbp]
        \centering
        \includegraphics[width=0.48\linewidth]{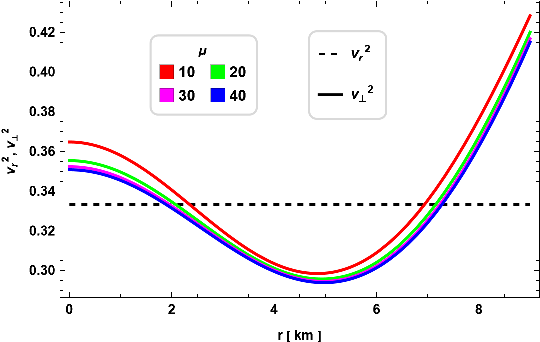}\,\,\,\,
        \includegraphics[width=0.48\linewidth]{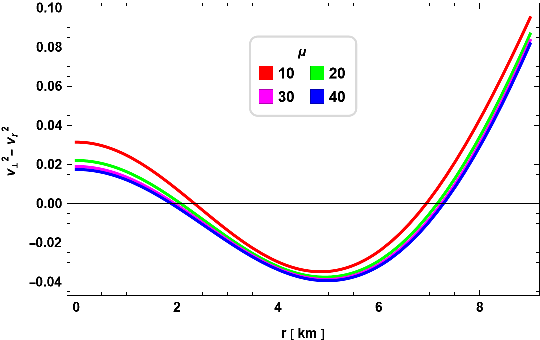}
        \caption{Left: Radial variation of the square of the radial velocity ($v_r^2$) and tangential velocity ($v_{\perp}^2$), and Right: Radial variation of ($|v_{\perp}^2 - v_r^2|$) for the compact object EXO 1745-248 ($M = 1.7\, M_{\odot}$ and $R = 9\, km$)}
        \label{fig:velocity}
    \end{figure}

For physical acceptability of the stellar model, it must satisfy the causality condition of the sound speed, which states that the square of the radial sound speed ($v_r^2 = \frac{dP_r}{d\rho} $) and square of the tangential sound speed ($v_\perp^2  = \frac{dP_\perp}{d\rho}$) must be within the range \cite{Herrera1992,Abreu2007}: 
    \begin{equation}
        0 \leq v_r^2 \left( = \frac{dP_r}{d\rho} \right) < 1 \quad \text{and} \quad 0 \leq v_\perp^2 \left(  = \frac{dP_\perp}{d\rho} \right) < 1.
    \end{equation}
The radial variation of $v_r^2$ and $v_\perp^2$, plotted in left sub-figure of Fig. (\ref{fig:velocity}), indicates that the quantities remain within the causal limit inside the star for the different values of the spheroidicity parameter $\mu$.  It is also clear from the figure that $v_r^2 = \frac{1}{3}$ throughoutthe star for all values of $\mu$ considered. The square of the tangential sound speed, however, attains a minimum near the boundary of the star and thereafter, attains a maximum value at the boundary of the star ($r=R$). 

The stability of the  stellar model is also analyzed by utilizing Abreu's condition \cite{Abreu2007} against Herrerra's cracking criteria. Herrera \cite{Herrera1992} claims that the term "cracking" in a fluid sphere of a self-gravitating compact object refers to the total radial forces (each with a different sign) that are created when any kind of perturbations are added to the system. Utilizing this cracking concept, Abreu demonstrated that the difference of the squares of the tangential and radial sound speeds can be employed to identify potentially unstable regions within a stellar configuration, i.e., $0 \leq |v_{\perp}^{2} - v_{r}^{2}| \leq 1$. The Abreu's index is plotted in the right sub-figure of Fig. (\ref{fig:velocity}) for different values of the spheroidicity parameter. It is found that Abreu's condition is within the limit and the stellar model is stable. 

\begin{figure}[tbp]
        \centering
        \includegraphics[height= 55 mm, width=0.48\linewidth]{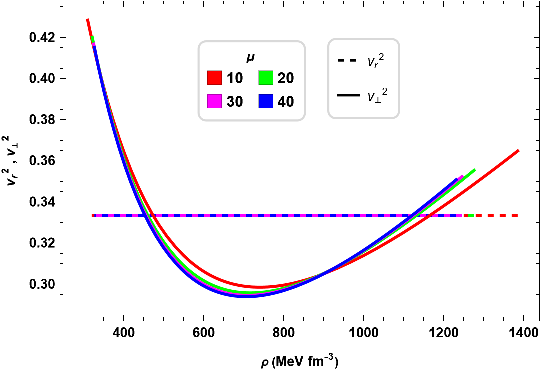}
        \caption{Variation of square of the speed of sound against energy density $\rho$ (MeV fm$^3$) for EXO 1745-248 ($M = 1.7\, M_{\odot}$ and $R = 9\, km$). The dotted line at $v_i^2 = 1/3$ indicates the conformal speed of sound.}
        \label{fig:velocity2}
     \end{figure}

The square of the radial and tangential sound speed are plotted against the energy density ($\rho$) in Fig. (\ref{fig:velocity2}) for a range of the spheroidicity parameter $\mu$. We find that for all values of the spheroidicity parameter considered, the square of radial sound speed is conformal ($v_r^2 = 1/3$) at any given density inside the star. However, we find a steep rise in the propagation of sound ($v_{\perp}^2 \gtrsim 1/3$) in the tangential direction at the lower densities ($\lesssim  450 MeV\, fm^{-3}$ ), which indicates a stiffening of the equation of state (EoS) at these densities for all values of $\mu$ \cite{Altiparmak2022}. However, a subsequent decrease in the sound propagation in the tangential direction is noted for $\rho \gtrsim 450 MeV\, fm^{-3}$ until $\rho \sim 650 MeV\, fm^{-3}$, where there is a minimum and the sound speed is subconformal ($v_r^2 < 1/3$). This indicates a softening of the EoS at these densities. Thereafter the sound speed increases again at higher density regions towards the center and becomes superconformal, i.e. $v_r^2 > 1/3$. This non-monotonic behavior of sound propagation in neutron stars is quite normal and this indicates that the EoS is relatively stiff towards the boundary of the star, where the sound propagation is higher than that of the core, where the EoS is less stiff. Between the core and the surface of the star, the EoS is found to be soft due to the presence of subconformal minima in sound speed \cite{Ecker2022}.

%%%%%%%%%%%%%%%%%%%%%%%%%%%%%%%%%%%%%%%%%%%%%%%%%%%%%%%%%%%%%%%%%%%%

\subsection{Adiabatic index}
\begin{figure}[tbp]
        \centering
        \includegraphics[width=0.5\linewidth]{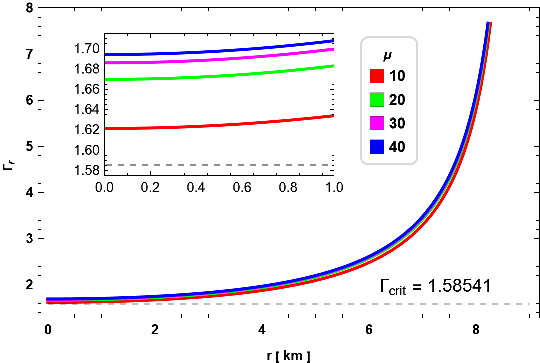}
        \caption{Radial variation of the adiabatic index ($\Gamma$) for the compact object EXO 1745-248 ($M = 1.7\, M_{\odot}$ and $R = 9\, km$)}
        \label{fig:adiabatic}
    \end{figure}

The analytical expression of the adiabatic index ($\Gamma$) is given by,
    \begin{equation}
         \Gamma = \left(1 + \frac{\rho}{P_r} \right) \frac{dP_r}{d\rho}.
    \end{equation}
Adiabatic index ($\Gamma$) indicates the stiffness of the EoS for a given energy density. In order to be stable against the Newtonian perturbations, the collapsing condition for a compact star with isotropic fluid distribution suggests that $\Gamma > \frac{4}{3}$. However, the collapsing condition changes in the relativistic limit \cite{Chan1992,Chan1993}, i.e.,
    \begin{equation}
 	\label{adia1}
 	\Gamma < \frac{4}{3} + \Bigg[\frac{1}{3}\kappa \frac{\rho_0 P_{r0}}{|P_{r0}^{'}|} + \frac{4}{3} \frac{(P_{\perp0} - P_{r0})}{|P_{r0}^{'}|r}  \Bigg]_{max},
    \end{equation}
where $\rho_0$, $P_{r0}$ and $P_{\perp0}$ are the initial density, radial pressure, and tangential pressure of the fluid at static equilibrium. The second and third terms on the right-hand side denote the relativistic and anisotropic corrections, respectively. Chandrasekhar \cite{Chandrasekhar1964a,Chandrasekhar1964b} suggested these relativistic modifications may cause instabilities within the stellar configuration. Moustakidis \cite{Moustakidis2017} proposed an additional constraint on the adiabatic index to address the instability issue, leading to a critical value of the adiabatic index, which is
    \begin{equation}
	\Gamma_{crit} = \frac{4}{3} + \frac{19}{21}u.
    \end{equation}
For a stable stellar configuration, the adiabatic index must be greater than this critical limit, which is determined by the compactness factor ($u$). It is clear from Fig. (\ref{fig:adiabatic}) that the adiabatic index is greater than the critical value ($\Gamma_{crit} = 1.58541$ for EXO 1745-248) throughout the star for the entire range of the spheroidicity parameter $\mu$. We also note that the adiabatic index at the center is increasing with an increase in $\mu$.

%%%%%%%%%%%%%%%%%%%%%%%%%%%%%%%%%%%%%%%%%%%%%%%%%%%%%%%%%%%%%%%%%%%%

\subsection{Zel'dovich condition}
\begin{figure}[tbp]
        \centering
        \includegraphics[width=0.48\linewidth]{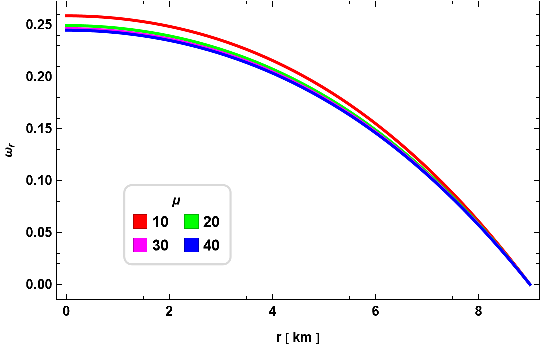}\,\,\,\,
        \includegraphics[width=0.48\linewidth]{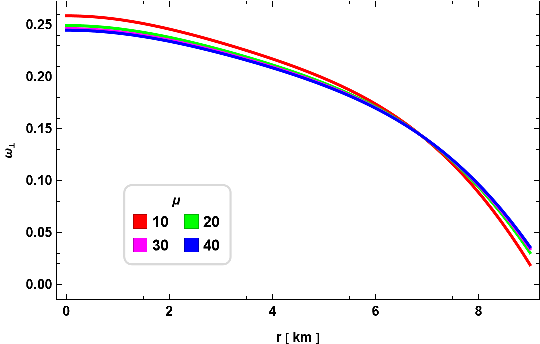}
        \caption{Radial variation of Pressure-to-energy density ratio ($\omega_r$ and $\omega_{\perp}$)  for the compact object EXO 1745-248 ($M = 1.7\, M_{\odot}$ and $R = 9\, km$)}
        \label{fig:zeldovich}
    \end{figure}

For a stable stellar configuration, the Zel'dovich condition states that the pressure-to-energy density ratio must be positive and less than unity inside the stellar structure \cite{Zeldovich1971}, i.e., $0 < \omega_r ( = \frac{P_{r}}{\rho}) < 1$ and $0 <  \omega_{\perp} ( = \frac{P_{\perp}}{\rho}) < 1$. It is clear from Fig. (\ref{fig:zeldovich}), that the Zel'dovich condition is satisfied for both the radial and tangential components of pressure, inside the stellar configuration.

    \begin{table}[h]
        \centering
        \begin{tabular}{c c c c c c}
            \hline
            $\mu$ &   $v_r^2(0)$ &  $v_{\perp}^2(0)$  &  $\Gamma(0)$  &  $\omega_r(0)$  & $\omega_{\perp}(0)$ \\
            \hline
             10   &   0.333     &   0.364804    &   1.62143 &   0.25878   &   0.25878\\
             20   &   0.3333    &   0.355412    &   1.66983 &   0.24941   &   0.24941\\
             30   &   0.3333    &   0.352342    &   1.68613 &   0.24640   &   0.24640\\
             40   &   0.3333    &   0.350818    &   1.69431 &   0.24492   &   0.24492\\
            \hline
        \end{tabular}
        \caption{Tabulation of physical parameters at the center ($r=0 $), required to analyze the stability, for different values of spheroidicity parameter ($mu$) for compact object EXO 1745-248 ($M = 1.7\, M_{\odot}$ and $R = 9\, km$)}
        \label{tab:stability}
    \end{table}
     
%%%%%%%%%%%%%%%%%%%%%%%%%%%%%%%%%%%%%%%%%%%%%%%%%%%%%%%%%%%%%%%%%%%%

\subsection{Hydrostatic Equilibrium}
\begin{figure}[tbp]
        \centering
        \includegraphics[width=0.6\linewidth]{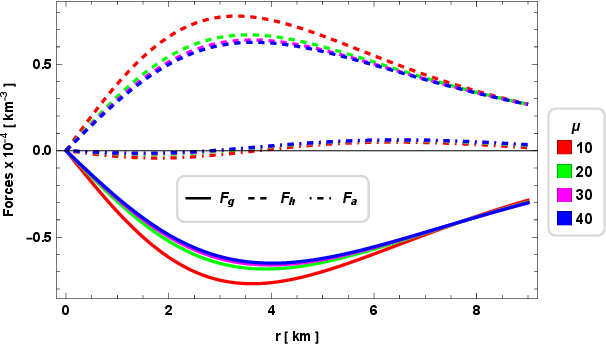}
        \caption{Hydrostatic equilibrium inside the compact object EXO 1745-248 ($M = 1.7\, M_{\odot}$ and $R = 9\, km$)}
        \label{fig:forces}
    \end{figure}

The Tolman-Oppenheimer-Volkoff (TOV) equation governs a spherically symmetric, isotropic body in a state of static gravitational equilibrium \cite{Tolman1939,Oppenheimer1939}. For an isotropic stellar body the TOV equation consists only the hydrostatic force gradient ($F_h$) and the gravitational force gradient ($F_g$). In the context of an anisotropic star within $f(Q)$ gravity, the TOV equation is supplemented by two additional force gradients, i.e., the anisotropic force gradient ($F_a$) and the force gradient arising from $f(Q)$ gravity ($F_Q$). The force gradients are defined as follows,
    \begin{equation}
        F_g = -\frac{\nu'(r)}{2}(\rho + P_{r}),
    \end{equation}
    \begin{equation}
 	F_h = -\frac{dP_{r}}{dr},
    \end{equation}
    \begin{equation}
 	F_a = \frac{2}{r} (P_{\perp} - P_{r}),
    \end{equation}
    \begin{multline}
        F_Q = \frac{e^{-\lambda}}{16 \pi  r^2} \Bigg[ -e^{\lambda } \Big(r^2 f'+ f_Q' \left(r \left(\lambda '+\nu '\right)+2\right)+r  \left(f_Q \left(\lambda ''+\nu ''\right)+2 f_{QQ} Q''\right)-f_Q \left(\lambda '+\nu '\right) \\ 
        +2 Q' \left(r f_{QQ}'+f_{QQ}\right)\Big) + r \, f_Q'\Big(  \lambda '+3  \nu ' \Big)- f_Q \Big(\lambda' + \nu' - r \lambda '' + r \lambda ' \nu ' + r \left(\lambda'\right)^2 - r \nu '' \Big) +2 r f_{QQ}' Q' \\
        + 2 f_{QQ} \Big(   r Q''-  r \lambda ' Q'-  r \nu ' Q'+ 2 Q' \Big) \Bigg] .
    \end{multline} 
Here, $f \equiv f(Q)$, $f_{Q} \equiv \frac{df(Q)}{dQ}$, and $f_{QQ} \equiv \frac{df_Q}{dQ}$ and prime ($'$) denote derivative with respect to $r$. For a linear form of $f(Q)$, the force gradient $F_Q$ vanishes and the modified TOV equation then consists of three terms only, i.e., $F_h$, $F_g$, and $F_a$. Therefore, the modified TOV equation takes the form:
\begin{equation}
	\label{tov}
	-\frac{\nu'}{2}(\rho+P_{r}) -\frac{dP_{r}}{dr} +  \frac{2}{r} (P_{\perp} -  P_{r})  = 0
\end{equation}
which can be expressed as,
	\begin{equation}
		\label{tov1}
		F_g + F_h + F_a = 0
	\end{equation}
The force gradients are plotted in Fig. (\ref{fig:forces}) for different values of the spheroidicity parameter $\mu$. It is evident that the combined forces within the star are maintaining the hydrostatic equilibrium as described by Eq. (\ref{tov1}). The anisotropy force gradient is negative towards the center of the star, i.e., attractive in nature, whereas towards the boundary it is positive, i.e., repulsive in nature. $F_g$ is negative, i.e., attractive in nature throughout the star and $F_h$ is positive, i.e., repulsive in nature throughout the stellar interior. The combination of these three forces thus maintains the hydrostatic equilibrium inside the stellar configuration.

%%%%%%%%%%%%%%%%%%%%%%%%%%%%%%%%%%%%%%%%%%%%%%%%%%%%%%%%%%%%%%%%%%%%

\section{Analyzing other strange star candidates}
\label{sec:9}
The admissibility of the stellar model of a strange star is also tested for three other strange star candidates, namely, PSR J1903+0327 ($ M = 1.667\,\pm 0.021\; M_\odot$  \& $R = 9.438\,\pm 0.03$ km) \cite{Freire2011OnJ1903+0327}, PSR J1614-2230 ($ M = 1.97\; M_\odot$  \& $R = 11$ km) \cite{Demorest2010ADelay}, and 4U 1608-52 ($ M = 1.74\,\pm 0.14\; M_\odot$  \& $R = 9.3\,\pm 1.0$ km) \cite{Guver2010THE160852}. Using the constraints on $\alpha$, the Mass-Radius relation and other acceptability criteria discussed in section \ref{sec:6}, we have found the lower limit on the spheroidicity parameter ($\mu$) for the said SS candidates for constructing a viable and physically acceptable stellar model. We note the following limits: for PSR J1903+0327 $\mu \geq 17$, for PSR J1614-2230 $\mu \geq 13$, and for 4U 1608-52 $\mu \geq 7$. In Tables \ref{tab:ss1} - \ref{tab:ss3}, the model parameters are tabulated along with the central and surface values of energy density, central pressure, causality analysis, and adiabatic index analysis for physically permitted values of the spheroidicity parameter ($\mu$). We further note that the model is only viable and physically acceptable for the SS candidates with a very high compactness factor ($u = \frac{M}{R} \geq 0.25$).

    \begin{table*}
        \caption{Tabulation of model parameters, central and surface values of energy density, central pressure, causality analysis, and adiabatic index analysis for SS candidate  {\bf PSR J1903+0327} \cite{Freire2011OnJ1903+0327} ($ M = 1.667\,\pm 0.021\; M_\odot$  \& $R = 9.438\,\pm 0.03$ km). Here we have considered $\phi = 2.036 \times 10^{-35}$, and $\alpha = 0.9$. The critical value of the adiabatic index, $\Gamma_{cric} = 1.56905$.}
        \label{tab:ss1}
        \begin{tabular}{c c c c c c c c c}
        \hline
     	  $\mu$ & $L$ & $D$ & $B_g$ & $\rho_c (\times 10^{14})$ & $\rho_R (\times 10^{14})$  & $P_c (\times 10^{14})$ & $0 \leq v_r^2,v_{\perp}^2 < 1$ & $\Gamma > \Gamma_{cric}$ \\
    	   & $(km^{-2})$  &   &  $(MeV/fm^{3})$ & $(gm/cm^{3})$ & $(gm/cm^{3})$ & $(gm/cm^{3})$ &  &\\
        \hline
        17	 &	 34.943	 &   0.33073 &	 74.920	 &  18.957  &   5.339   &   4.539   & satisfied & satisfied \\
        20	 &	 38.297	 &   0.33161 &	 75.327	 &  18.741  &   5.368   &   4.458   &   satisfied & satisfied \\
        30	 &   47.806  &   0.33319 &   76.076  &  18.357  &   5.421   &   4.312   &   satisfied & satisfied \\
        40	 &   55.716  &   0.33396 &   76.441  &  18.175  &   5.447   &   4.243   &   satisfied & satisfied \\
        \hline
        \end{tabular}

        \caption{Tabulation of model parameters, central and surface values of energy density, central pressure, causality analysis, and adiabatic index analysis for SS candidate  {\bf PSR J1614-2230} \cite{Demorest2010ADelay} ($ M = 1.97\; M_\odot$  \& $R = 11$ km). Here we have considered $\phi = 2.036 \times 10^{-35}$, and $\alpha = 1.3$. The critical value of the adiabatic index, $\Gamma_{cric} = 1.57233$.}
        \label{tab:ss2}
        \begin{tabular}{c c c c c c c c c} 
        \hline
     	  $\mu$ & $L$ & $D$ & $B_g$ & $\rho_c (\times 10^{14})$ & $\rho_R (\times 10^{14})$  & $P_c (\times 10^{14})$ & $0 \leq v_r^2,v_{\perp}^2 < 1$ & $\Gamma > \Gamma_{cric}$ \\
    	   & $(km^{-2})$  &   &  $(MeV/fm^{3})$ & $(gm/cm^{3})$ & $(gm/cm^{3})$ & $(gm/cm^{3})$ &  &\\
        \hline
        13	 &	 34.283	 &   0.31991 &	 79.179	 &  21.334  &   5.642   &   5.231   & satisfied & satisfied \\
        20	 &	 43.949	 &   0.32274 &	 80.564	 &  20.555  &   5.741   &   4.938   &   satisfied & satisfied \\
        30	 &   54.881  &   0.32437 &   81.382  &  20.120  &   5.799   &   4.774   &   satisfied & satisfied \\
        40	 &   63.969  &   0.32516 &   81.781  &  19.915  &   5.828   &   4.696   &   satisfied & satisfied \\
        \hline
        \end{tabular}

        \caption{Tabulation of model parameters, central and surface values of energy density, central pressure, causality analysis, and adiabatic index analysis for SS candidate  {\bf 4U 1608-52} \cite{Guver2010THE160852} ($ M = 1.74\,\pm 0.14\; M_\odot$  \& $R = 9.3\,\pm 1.0$ km). Here we have considered $\phi = 2.036 \times 10^{-35}$, and $\alpha = 0.9$. The critical value of the adiabatic index, $\Gamma_{cric} = 1.58302$.}
        \label{tab:ss3}
        \begin{tabular}{c c c c c c c c c}
        \hline
     	  $\mu$ & $L$ & $D$ & $B_g$ & $\rho_c (\times 10^{14})$ & $\rho_R  (\times 10^{14})$  & $P_c (\times 10^{14})$ & $0 \leq v_r^2,v_{\perp}^2 < 1$ & $\Gamma > \Gamma_{cric}$ \\
    	   & $(km^{-2})$  &   &  $(MeV/fm^{3})$ & $(gm/cm^{3})$ & $(gm/cm^{3})$ & $(gm/cm^{3})$ &  &\\
        \hline
        10	 &	 23.354	 &   0.28825 &	 76.583	 &  23.871  &   5.457   &   6.138   & satisfied & satisfied \\
        20	 &	 35.321	 &   0.29423 &	 79.370	 &  22.032  &   5.656   &   5.459   &   satisfied & satisfied \\
        30	 &   44.155  &   0.29602 &   80.235  &  21.518  &   5.718   &   5.267   &   satisfied & satisfied \\
        40	 &   51.496  &   0.29687 &   80.657  &  21.276  &   5.748   &   5.176   &   satisfied & satisfied \\
        \hline
        \end{tabular}

%% Any table notes must follow the \end{tabular} command.
%\tablenotetext{a}{Sample footnote for table~\ref{tbl-2} that was
%generated with the \LaTeX\ table environment}
%\tablenotetext{b}{Yet another sample footnote for table~\ref{tbl-2}}
%\tablenotetext{c}{Another sample footnote for table~\ref{tbl-2}}
%\tablecomments{We can also attach a long-ish paragraph of explanatory
%material to a table}
    \end{table*}

%%%%%%%%%%%%%%%%%%%%%%%%%%%%%%%%%%%%%%%%%%%%%%%%%%%%%%%%%%%%%%%%%%%%

\section{Discussion }
\label{sec:10}
In the paper, we obtain singularity-free relativistic interior solutions in the framework of a linear form of $f(Q)$ gravity ($f(Q) = \alpha Q + \phi$) for constructing stable stellar models and have explored physical features along with the stability of an SS candidate, namely, EXO 1745-248. In the metric, the spatial coefficient ($e^{\lambda(r)}$) of the line element is described by the pseudo-spheroidal geometry, and the interior composition is given by the MIT Bag model EoS. The metric potential $e^{\nu(r)}$ is obtained making use of the field equations in $f(Q)$ gravity employing the Durgapal-Banerjee transformation \cite{Durgapal1983NewRelativity}. For a linear form of $f(Q)$ gravity, we obtain the exterior vacuum solution, which reduces to the Schwarzschild-de Sitter (SdS) solution with the cosmological constant term, $\Lambda = \frac{\phi}{2\alpha}$. Following the work of Chattopadhyay et.al. \cite{Chattopadhyay2010RelativisticSpace-time}, we analyze the stellar models for the given values of the spheroidicity parameter: $\mu = 10,\, 20,\, 30,\, \&\,  40$. We constrain the value of $\alpha$ using a viable physical limit on the Bag constant ($B_{g}$) of the MIT Bag model equation of state (EoS). The contribution of the spheroidicity parameter to the energy density, pressure profiles, and other physical features is studied for EXO 1745-248. The stability of the stellar models obtained here is also analyzed. We note the following:\\

(i) The range of permissible values of $\alpha$ is determined with a viable limit of the Bag constant ($B_{g}$). A contour plot of the physical limit on the Bag constant ($B_{g} \in [57.55,95.11]$ $ MeV\,fm^{-3}$) is plotted in Fig. (\ref{fig:alphaconstraint}) for different values of the spheroidicity parameter $\mu$, which shows the permitted values of $\alpha$ for constructing a stable stellar model. We choose $\alpha = 0.85$ to study the physical features of the stellar model.

(ii) The Mass-Radius (M-R) relation for the stellar model is constrained by the latest observable constraints in Fig. (\ref{fig:MRconstraints}) for different values of $\mu$. The latest stringent constraints on the mass-radius relation of neutron stars are determined by the maximum mass, minimum radius, highest rotational frequency, and maximum surface gravity observed in pulsars \citep{Trumper2011ObservationsDensities}. It is found that only for $\mu \geq 7$, we get allowed M-R relation, that is, passing through the area (gray area in Fig. (\ref{fig:MRconstraints})) delimited by the constraints. 

(iii) It is evident from Fig. (\ref{fig:metrics}), that the interior metric potentials ($e^{\nu(r)}$ and $e^{\lambda(r)}$) are finite at the center of the star and are matching with the Schwarzschild-de Sitter metric at the boundary of the star for a realistic stellar model. Moreover, the metric potentials are found to be regular inside the stellar interior.
        
(iv) The energy density ($\rho$) and both the radial and tangential pressure profiles ($P_r$ and $P_{\perp}$, respectively) are found to be finite and positive inside the stellar interior. It is clear from Figs. (\ref{fig:density}) and (\ref{fig:pressure}), that the energy density and the pressure profiles are maximum at the center of the star, whcich are monotonically decreasing away from the center. The gradients of energy density and pressure profiles are also found to be negative throughout the interior of the star, as evident from Figs. (\ref{fig:density}) and (\ref{fig:gradpressure}), respectively. Central and surface values of energy density and pressure value at the center for different values of the spheroidicity parameter are tabulated in Table. \ref{tab:physicalparameters}.

(v) The anisotropy ($\Delta$) profile is plotted in Fig. (\ref{fig:anisotropy}) for different spheroidicity parameter ($\mu$) values. The anisotropy vanishes at the center for any given value of $\mu$, which indicates the stability of the stellar model. As we move away from the center, $\Delta$ is negative up to a certain radial distance where $\Delta$ vanishes; thereafter, $\Delta$ begins to increase and attains a maximum near the surface of the star.

(vi) The radial variation of the energy conditions in Fig. (\ref{fig:energyconditions}) shows that all the energy conditions are satisfied for the values of the spheroidicity parameter ($\mu$) considered here.

(vii) The stability of the stellar model is ensured from the radial variation of the square of the sound speed ($v_r^2$ and $v_{\perp}^2$) in Fig. (\ref{fig:velocity}), which shows that the propagation of sound speed is causal in the stellar interior. Moreover, it is clear from the figure that Abreu's condition is satisfied for all the values of spheroidicity parameter ($\mu$) considered. The variation of the square of the sound speed against the energy density ($\rho$) in Fig. (\ref{fig:velocity2}) suggests that the square of the radial sound speed, $v_r^2 = 1/3$ i.e., is conformal at any given energy density for any value of $\alpha$. However, the square of the tangential sound speed ($v_{\perp}^2$) is superconformal, and it attains a maximum at the lower densities ($\lesssim 450\, MeV \, fm^{-3} $) towards the boundary of the star. For $\gtrsim 450\, MeV \, fm^{-3} $, there is a subsequent decrease in $v_{\perp}^2$ until $\sim 650\, MeV \, fm^{-3} $, where we note a minimum and the sound speed is subconformal ($v_{\perp}^2 < 1/3$). Thereafter, $v_{\perp}^2$ starts increasing again and becomes superconformal at very high density at the center of the star. This non-monotonic behavior inside the neutron star indicates that the equation of state (EoS) is relatively stiffer towards the boundary of the neutron star than that of the core, where the EoS is less stiff. Between the core and the surface, the EoS is found to be soft due to the subconformal sound speed.

(vii) The radial variation of the adiabatic index ($\Gamma$) plotted in Fig. (\ref{fig:adiabatic}) shows that $\Gamma$ is greater than the critical value ($\Gamma_{crit} = 1.58541$) at the center of the star for all values of $\mu$ considered. We also note that $\Gamma$ is increasing at the center of the star with an increase in $\mu$.

(viii) In Fig. (\ref{fig:zeldovich}),  the pressure-to-energy density ($\omega_r$ and $\omega_{\perp}$) ratios are plotted for different values of $\mu$. It is evident that the Zel'dovich condition is satisfied inside the stellar configuration.

(ix) The radial variation of the force gradients are plotted in Fig. (\ref{fig:forces}) for different values of $\mu$. We found that the gravitational force gradient ($F_g$) is negative, i.e., attractive in nature, and the hydrostatic force gradient ($F_h$) is positive, i.e., repulsive in nature. The anisotropic force gradient ($F_a$) is negative towards the center but becomes positive away from the center of the star. The combination of these three force gradients thus satisfies the hydrostatic equilibrium condition inside the stellar configuration.

(x) A few known strange star candidates, such as PSR J1903+0327, PSR J1614-2230, and 4U 1608-52 are also analyzed to test the admissibility of the model. For a physically acceptable stellar model we obtain different lower limits of the spheroidicity parameter for different stars with given mass and radius. We note that only the strange stars with a very high compactness factor ($u \geq 0.25$) are acceptable.

Therefore, we obtain a new class of singularity-free interior solution relevant for the description of a realistic anisotropic strange star model in the framework of linear $f(Q)$ gravity with the pseudo-spheroidal geometry. From the analysis of the constraints on the Mass-Radius relation, it can be concluded that for any given star with $\mu \geq 7$ the model satisfies several physical features and stability criteria, which are required for a physically viable stellar configuration. The implications of the results obtained here, in relation to current observational data of relativistic compact stars, need to be investigated further in other modified theories of gravity and will be addressed in a separate study.

\acknowledgments
BD is thankful to CSIR, New Delhi for financial support. The authors would like to thank  IUCAA Centre for Astronomy Research and Development (ICARD), NBU for the research facilities.

\bibliographystyle{ieeetr} 
\bibliography{f(Q)_Linear}

\end{document}